\documentstyle{mn}
\begin{document}

\def\kms{${\rm km\,s}^{-1}$}
\def\HII{H\,{\sc ii}}
\def\etal{et~al.}
\def\methanol{$\rm{CH_{3}OH}$}
\def\6ghz{$5_{1}-6_{0}\,A^{+}$}
\def\12ghz{$2_{0}-3_{-1}\,E$}
\def\cc{$\,\rm cm^{-3}$}
\def\ccs{$\,\rm cm^{-3} s$}
\def\WHII{\mbox{$W_{\rm HII}$}}
\def\nH{\mbox{$n_{\rm H}$}}
\def\scd{\mbox{$N_{\rm M}/\Delta V$}}
\def\Td{\mbox{$T_{d}$}}
\def\Tk{\mbox{$T_{k}$}}
\def\wn{$\rm cm^{-1}$}
\def\vt{\mbox{$v_{t}$}}
\def\arcmin{\hbox{$^\prime$}}
\def\arcsec{\hbox{$^{\prime\prime}$}}
\def\degr{\hbox{$^{\circ}$}}
\def\beam{\mbox{$\epsilon^{-1}$}}
\def\Tb{\mbox{$T_{b}$}}

\input epsf
\epsfverbosetrue

\title[Multi-transition study and new detections of class II methanol
masers]{Multi-transition study and new detections of class II methanol masers}

\author[D.M. Cragg et al.]{Dinah M. Cragg $^1$,
 Andrej M. Sobolev $^2$,
 Simon P. Ellingsen $^3$,
 J. L. Caswell $^4$,
\newauthor Peter D. Godfrey $^1$,
 Svetlana V. Salii $^2$, 
 and Richard G. Dodson $^3$\\
$^1$ Department of Chemistry, Monash University, Clayton, Victoria 3800, Australia;  Dinah.Cragg@sci.monash.edu.au,\\  Peter.Godfrey@sci.monash.edu.au\\
$^2$ Astronomical Observatory, Ural State University, Lenin Street 51, Ekaterinburg 620083, Russia;  Andrej.Sobolev@usu.ru,\\
Svetlana.Salii@usu.ru\\
$^3$ School of Mathematics and Physics, University of Tasmania, GPO Box 252-21, Hobart, Tasmania 7001, Australia;\\  Simon.Ellingsen@utas.edu.au\\
$^4$ Australia Telescope National Facility, CSIRO, PO Box 76, Epping, NSW 2121, Australia;  James.Caswell@atnf.csiro.au\\}

\maketitle

\begin{abstract}

We have used the ATNF Mopra antenna and the SEST antenna to search in the directions of several class~II methanol maser sources for emission from six methanol transitions in the frequency range 85-115~GHz.  The transitions were selected from excitation studies as potential maser candidates.  Methanol emission at one or more frequencies was detected from five of the maser sources, as well as from Orion~KL.  Although the lines are weak, we find evidence of maser origin for three new lines in G345.01+1.79, and possibly one new line in G9.62+0.20. 

The observations, together with published maser observations at
other frequencies, are compared with methanol maser modelling for
G345.01+1.79 and NGC~6334F.  We find that the majority of observations in both sources are consistent with a  warm dust (175~K) pumping model at hydrogen density $\sim 10^{6}$~\cc\ and methanol column density $\sim 5 \times 10^{17} \,\rm cm^{-2}$.  The substantial differences between the maser spectra in the two sources can be attributed to the geometry of the maser region.

\end{abstract}

\begin{keywords}
\HII\ regions --- ISM: molecules --- masers --- radio lines : ISM
\end{keywords}

\section{Introduction}

Class~II methanol masers provide a unique high resolution tracer of physical conditions at the sites of massive star formation.  The two strongest class~II methanol maser transitions are the \6ghz\/ transition at 6.6~GHz and the \12ghz\/ transition at 12.1~GHz.  Single dish and interferometric studies of these transitions (e.g. Menten 1991, Caswell \etal\ 1995a,b, Norris \etal\ 1993) have revealed the widespread distribution of the masers, the forms of maser spot distributions, and their velocity fields.  However, data on only two transitions are not sufficient to determine physical parameters in the sources through modelling because there are so many model parameters (see, e.g., Sobolev, Cragg \& Godfrey 1997a).  

In addition to the very bright 6.6- and 12.1-GHz methanol masers,
moderately bright class~II masers have been identified at 19.9~GHz (Wilson
\etal\ 1985), 23.1~GHz (Wilson \etal\ 1984), 28.9~GHz (Wilson \etal\
1993), 37.7, 38.2 and 38.4~GHz (Haschick, Baan \& Menten 1989), 107.0~GHz
(Val'tts \etal\ 1995), 156.6~GHz, and in a series of eight lines near
157~GHz (Slysh, Kalenskii \& Val'tts 1995).  More recently a new weaker
class~II maser line has been identified at 108.8~GHz (Val'tts \etal\
1999).

A model featuring pumping by IR radiation to the second and first
torsionally excited states of methanol, together with amplification of
background \HII\ region continuum radiation, can account for the extreme
brightness of the 6.6- and 12.1-GHz masers (Sobolev \& Deguchi 1994a,
Sobolev \etal\ 1997a).  Furthermore, it predicts maser action spanning a
wide range of brightness temperatures in $\sim 100$ other methanol lines
(Sobolev, Cragg \& Godfrey 1997b, hereafter SCG97).  This list of maser
candidates includes all 18 established class~II masers and also the more
recently detected 108.8-GHz weak maser.

Here, we extend the search for weaker class~II methanol maser
lines, based on the list of maser candidate transitions from
SCG97.  Sections 2 and 3 report a search for the five remaining
transitions in the 85-115~GHz band not previously observed in class~II
maser sources.  We also searched for a line of the first torsionally
excited state of methanol (\vt=1), predicted to have class~II
maser character, but too weak to be included by 
SCG97.  The quantum numbers and rest frequencies of the six lines are
given in Table~\ref{tab:freqs}.

Since the relative brightness of different methanol masers is sensitive to
the model parameters (Section 4.1), multi-transition studies provide a
probe of the physical conditions in star-forming regions if the masers are co-spatial (Section 4.2).  We compare the model with all available methanol
maser observations for G345.01+1.79 (Section 4.3) and NGC~6334F (Section 4.4).

\begin{table}

\caption{Methanol laboratory frequencies from Tsunekawa \etal\ (1995).  Figures in brackets represent uncertainty in kHz.  The last column is the velocity resolution (after Hanning smoothing) of the astronomical observations reported here.}

\label{tab:freqs}

\centerline{
\begin{tabular}{clrc} \hline
\vt & Transition & \multicolumn{1}{c}{Frequency} & \multicolumn{1}{c}{Resolution} \\
&& \multicolumn{1}{c}{(MHz)} & \multicolumn{1}{c}{(\kms)} \\ \hline
0 & $6_{-2}-7_{-1}\,E$ & 85568.084(30) & 0.28 \\
0 & $7_{2}-6_{3}\,A^{-}$ & 86615.578(30) & 0.43 \\
0 & $7_{2}-6_{3}\,A^{+}$ & 86902.956(50) & 0.43 \\
1 & $1_{0}-2_{1}\,E$ & 93196.682(30) & 0.40 \\
0 & $8_{3}-9_{2}\,E$ & 94541.778(50) & 0.40 \\
0 & $7_{2}-8_{1}\,A^{+}$ & 111289.515(50) & 0.34 \\ \hline
\end{tabular}}

\end{table}

\section{Observations}

The 85.5-GHz observations were made at SEST\footnote{SEST=Swedish-ESO Submillimetre Telescope, La Silla, Chile} 1996 March using standard SEST beam switching and calibration (http://www.ls.eso.org/lasilla/Telescopes/SEST).  The high resolution acousto-optical spectrometer provided a 2048-channel spectrum over an 86-MHz bandwidth centred near 85568~MHz.  The antenna half-power beam width (HPBW) at 85.5-GHz is 58\arcsec, the conversion efficiency for a point source is 25.2 Jy/K, and the pointing accuracy is better than 5\arcsec.  The system temperature was about 130~K, and signal integration periods were 12~min (exceptions were G9.62+0.20 -- 32~min and G12.91-0.26 -- 8~min).  The data were processed using \lq CLASS' from the GILDAS software package.   

The remaining observations were made with the Mopra
Antenna, operated by the Australia Telescope National Facility,
ATNF\footnote{The Australia Telescope is funded by the Commonwealth of
Australia for operation as a National Facility managed by
CSIRO.}.  Observations at 86.6, 86.9 and 93.1~GHz were made in 1996
November, and at 94.5 and 111.2~GHz in 1998 May.  The illuminated diameter
of the antenna was 15.4~m, with HPBW 37\arcsec\ at 111~GHz increasing to
48\arcsec\ at 86~GHz. 

Spectra from a digital correlator comprised two bands,
each of 1024 channels spanning 64~MHz.  One band was tuned to the methanol
transition, giving a channel separation of approximately 0.2 \kms\ 
(Table~\ref{tab:freqs}).  Generally the second
band was reserved for 86.2-GHz SiO maser pointing calibration, but the
86.6-GHz methanol observation was made simultaneously with the 86.9-GHz
observation by retuning this second band.  On-source observations were
interleaved with reference spectra (offset by 30\arcmin\ or 60\arcmin\ in
declination) using five-minute integration times.  

The 1996 November data at 86.6, 86.9 and 93.1~GHz were not systematically 
corrected for atmospheric attenuation; rather, the recorded
antenna temperatures have been scaled by a factor of 2, indicated from 
comparison between our 86-GHz Orion~KL spectra and those observed by
Johansson \etal\ (1984) at Onsala, and by Turner (1991) at Kitt
Peak, while taking into account the different beam sizes.  The intensity
calibration is uncertain to a factor of 2, but adequate for a detection
experiment, and for order-of-magnitude comparisons with modelling. We applied a conversion factor of 50 Jy/K to the corrected
antenna temperatures.  

During the 1998 May observations at 94.5 and 111.2~GHz, additional measurements of receiver total power were made with an ambient temperature load (assumed temperature 290~K) regularly inserted.  These alloow correction for atmospheric absorption using the method of Kutner \& Ulich (1981). The uncertainty of the absolute flux density scale of these observations is 20\%. 

\begin{table*}

\caption{Upper limits in Jy for methanol line search (3 $\times$ rms);  D denotes detection, with details in Table~\ref{tab:obs}.  Source coordinates are from Caswell \etal\ (1995a).}

\label{tab:limits}

\centerline{
\begin{tabular}{lllcccccc} \hline
Source & RA(1950) & Dec(1950) &  85.5~GHz & 86.6~GHz & 86.9~GHz & 93.1~GHz & 94.5~GHz & 111.2~GHz \\ 
& \multicolumn{1}{c}{(h m s)} & \multicolumn{1}{c}{(\degr\ \arcmin\ \arcsec)}  &&&&& \\ \hline
Orion KL     & 05 32 47.0   & -05 24 23.0  && D & D & D & D &   \\
G188.95+0.89 &  06 05 53.5  & +21 39 02    && $<$0.8 & $<$0.9 & $<$1.5 &&  \\
G305.21+0.21 &  13 08 01.72 &  -62 18 45.3 && $<$1.4 & $<$1.4 && $<$2.3 & $<$4.2 \\
G309.92+0.48 &  13 47 11.85 &  -61 20 18.8 && $<$1.9 & $<$1.6 &&& \\
G318.95-0.20 &  14 57 03.81 &  -58 47 01.2 && $<$2.5 & $<$2.8 &&& \\
G323.74-0.26 &  15 27 51.97 &  -56 20 39.5 && $<$5.1 & $<$2.7 & 8.0 & $<$3.8 & $<$6.1 \\
G328.81+0.63 &  15 52 00.32 &  -52 34 22.2 && $<$2.4 & $<$3.1 &&& \\
G336.02-0.83 &  16 31 24.8  &  -48 40 37   && $<$2.7 & $<$4.6 &&& \\
G339.88-1.26 &  16 48 24.76 &  -46 03 33.9 && $<$1.7 & $<$2.1 & $<$5.3 & $<$2.5 &  \\
G345.00-0.22 &  17 01 38.5  &  -41 24 59   && $<$2.9 & $<$2.7 &&& \\
G345.01+1.79 &  16 53 19.69 &  -40 09 46.0 & D & D & D & $<$3.1 & $<$2.5 &  \\
NGC 6334F    &  17 17 32.3  &  -35 44 04   & D & D & D & D & D & \\
G351.77-0.54 &  17 23 20.67 &  -36 06 45.4 && D & D &&& \\
G9.62+0.20   &  18 03 15.98 &  -20 31 52.9 & D & $<$2.3 & $<$2.2 & $<$2.3 & $<$2.5 &  \\
G10.47+0.03  &  18 05 40.0  &  -19 52 24   & D & $<$2.9 & $<$2.8 & $<$3.1 && \\
G12.68-0.18  &  18 10 59.6  &  -18 02 29   &&& $<$5.2 &&& \\
G12.91-0.26  &  18 11 43.8  &  -17 53 04   & $<$3.8 &&&&& \\
G23.01-0.41  &  18 31 55.6  &  -09 03 09   & $<$3.8 &&&&& \\
G31.28+0.06  &  18 45 37.2  &  -01 30 00   & $<$3.8 &&&&& \\
G35.20-1.74  &  18 59 13.1  &  +01 09 07   & $<$3.8 & $<$2.3 & $<$2.8 & $<$4.9 & $<$3.6 & \\ \hline
\end{tabular}}

\end{table*}

The data were processed using the SPC data reduction package at the 
ATNF.  After averaging individual spectra and subtracting baselines,
Hanning smoothing was applied and Gaussian line profiles fitted. Velocity tracking was not available at Mopra, but the LSR velocity shift over the timeframe of our observations was generally less than one channel, so that the spectra are not significantly smeared.  A frequency setting problem at 93.1~GHz led us to adjust the velocity scale by 1.61 \kms\ following comparison of the sharply defined spectrum of $\rm{N_{2}H^{+}}$ in L134N with previous observations (Womack, Ziurys \& Wyckoff 1992).

Class~II maser sources selected from the surveys of Caswell \etal\
(1995a,b) are listed in Table~\ref{tab:limits} and were observed at the
various methanol frequencies as listed in Table~\ref{tab:freqs}.  As a system test at Mopra we reobserved previously detected thermal methanol emission from Orion~KL.

\section{Results}

The present observations detected weak high frequency methanol emission from five of the class~II methanol maser sources, as well as Orion~KL.  The parameters of peak flux density, mean velocity, and width to half intensity, as derived from Gaussian fits to the observed methanol lines, are given in Table~\ref{tab:obs}.  Upper limits where no line was detected are summarised in Table~\ref{tab:limits}.  The spectra of detected methanol lines are presented in Figs.~1-4.

It is not always easy to distinguish between weak maser emission from a very compact source and thermal emission from an area that is more extended, but nevertheless smaller than a single dish telescope beam.  Many sources of class~II methanol maser emission also exhibit thermal methanol features at millimetre wavelengths (e.g. Caswell \etal\ 2000).  In identifying possible new masers we have been guided by the width of the line and the velocity range exhibited by previous maser and thermal observations.  In the sources examined all firmly established maser features have widths of the order 0.5~\kms, while the thermal features are considerably broader in the range $3-10$~\kms.  Only features narrower than 2~\kms\ appearing at the characteristic methanol maser velocity of the source are considered here as probable new masers (see discussion of sources in the following subsections).

\begin{table}

\caption{Summary of methanol line detections.  Figures in brackets are uncertainties from Gaussian line fitting.}

\label{tab:obs}

\centerline{
\begin{tabular}{lrrr} \hline
Source & \multicolumn{1}{c}{$S$} & \multicolumn{1}{c}{$V_{LSR}$} & \multicolumn{1}{c}{$\Delta V$}  \\ 
& \multicolumn{1}{c}{(Jy)} & \multicolumn{1}{c}{(\kms)} & \multicolumn{1}{c}{(\kms)} \\ \hline
\multicolumn{4}{l}{\bf{85.5~GHz}}  \\
G345.01+1.79 & 10.0(0.8) & -22.1(0.1) & 0.8(0.1)  \\
NGC 6334F & 7.7(1.0) & -8.3(0.1) & 5.4(0.1)  \\
G9.62+0.20 & 0.9(0.1) & 3.8(0.3) & 5.7(0.6)  \\
\multicolumn{1}{c}{\arcsec} & 1.2(0.4) & -0.9(0.1) & 0.5(0.1)  \\
G10.47+0.03 & 3.3(1.0) & 66.2(0.2) & 9.0(0.4)  \\
&&&  \\
\multicolumn{4}{l}{\bf{86.6~GHz}}  \\
Orion KL & 12.3(0.7) & 8.0(0.3) & 4.2(0.3)  \\
G345.01+1.79 & 2.8(0.3) & -21.9(0.3) & 1.5(0.2)  \\
NGC 6334F & 2.6(0.2) & -8.1(0.4) & 5.5(0.4)  \\
G351.77-0.54 & 1.5(0.2) & -3.1(0.6) & 9.8(1.0)   \\
&&&  \\
\multicolumn{4}{l}{\bf{86.9~GHz}}  \\
Orion KL & 14.2(0.3) & 8.1(0.3) & 4.4(0.1)  \\
G345.01+1.79 & 4.1(0.4) & -21.7(0.3) & 1.0(0.2)  \\
NGC 6334F & 3.6(0.2) & -8.0(0.3) & 4.6(0.3) \\
G351.77-0.54 & 1.8(0.2) & -3.5(0.8) & 9.3(1.4)  \\
&&&  \\
\multicolumn{4}{l}{\bf{93.1~GHz}}  \\
Orion KL & 3.5(0.3) & 7.6(0.3) & 3.2(0.3)  \\
NGC 6334F & 3.4(0.2) & -8.3(0.4) & 3.6(0.4)  \\
&&&  \\
\multicolumn{4}{l}{\bf{94.5~GHz}}  \\
Orion KL & 14.7(0.4) & 7.6(0.1) & 4.5(0.1)  \\
NGC 6334F & 4.4(0.4) & -7.9(0.3) & 5.0(0.6)  \\ \hline
\end{tabular}}

\end{table}

\begin{figure}
\epsfxsize=8cm\epsfbox{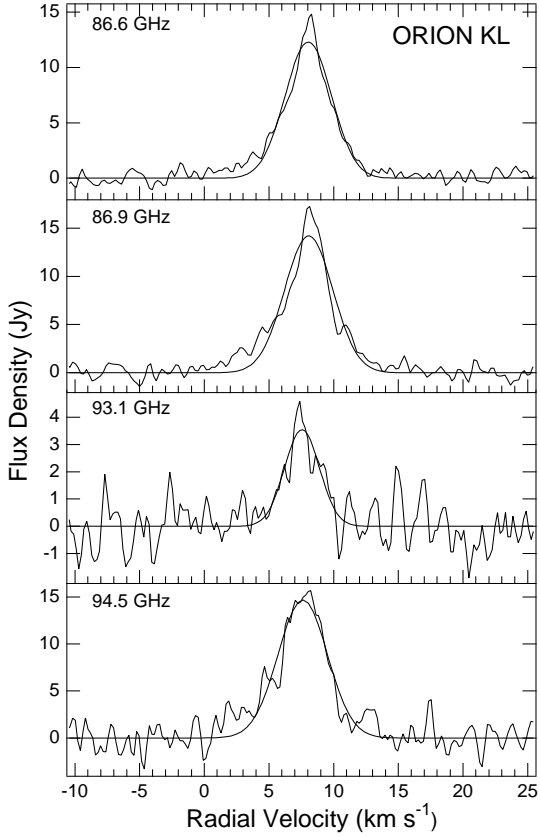}
\caption{Methanol spectra observed at Mopra for Orion KL. }
\label{fig:orion}
\end{figure}

\begin{figure}
\epsfxsize=8cm\epsfbox{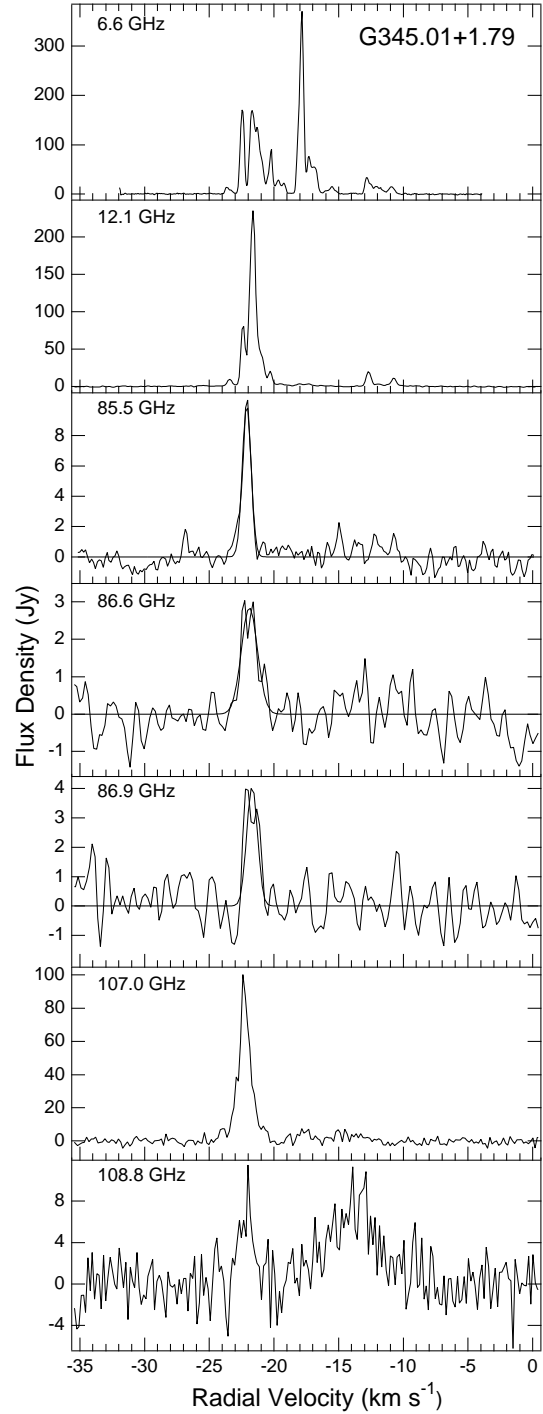}
\caption{Methanol spectra observed at SEST (85.5 GHz) and Mopra (86.6,
86.9 GHz) for G345.01+1.79.  For comparison we include spectra at 107.0
and 108.8 GHz from Val'tts \etal\ (1999), and at 6.6 and 12.1~GHz 
from unpublished observations made with the University of Tasmania 26m Mt
Pleasant antenna.}
\label{fig:g345}
\end{figure}

\begin{figure}
\epsfxsize=8cm\epsfbox{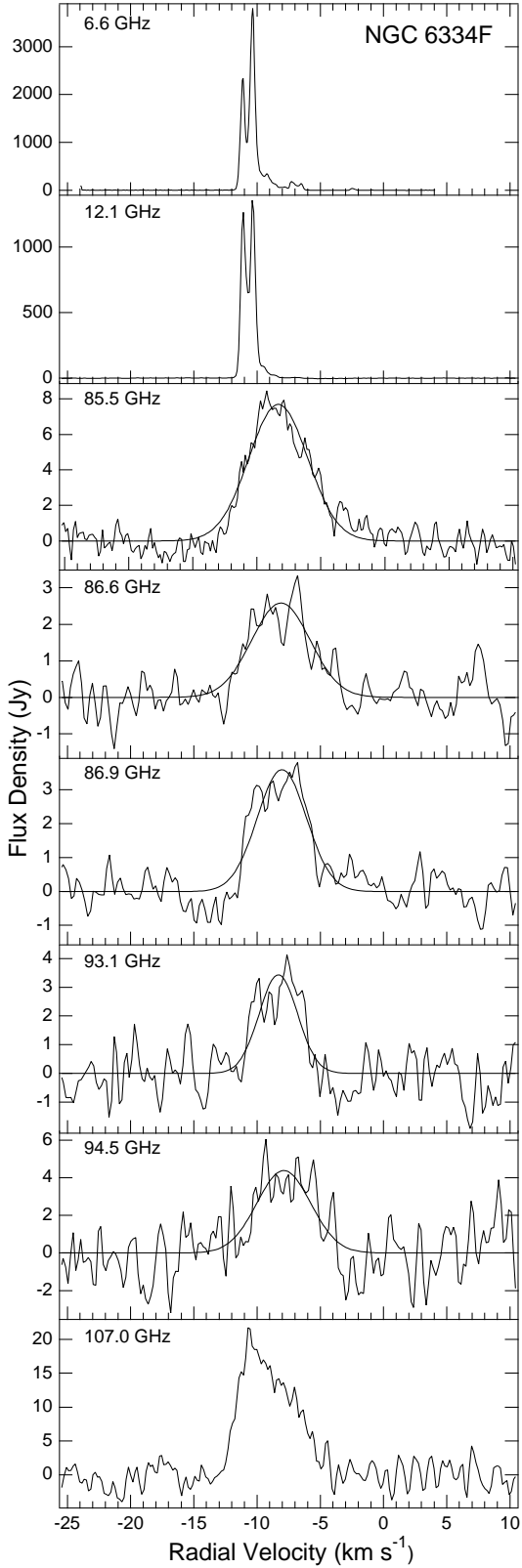}
\caption{Methanol spectra observed at SEST (85.5~GHz) and Mopra (86.6,
86.9, 93.1,94.5~GHz) for NGC6334F, with additional spectra at 107.0 GHz,
6.6 GHz, and 12.1 GHz, as in Fig.~\ref{fig:g345}. }
\label{fig:ngc}
\end{figure}

\begin{figure}
\epsfxsize=8cm\epsfbox{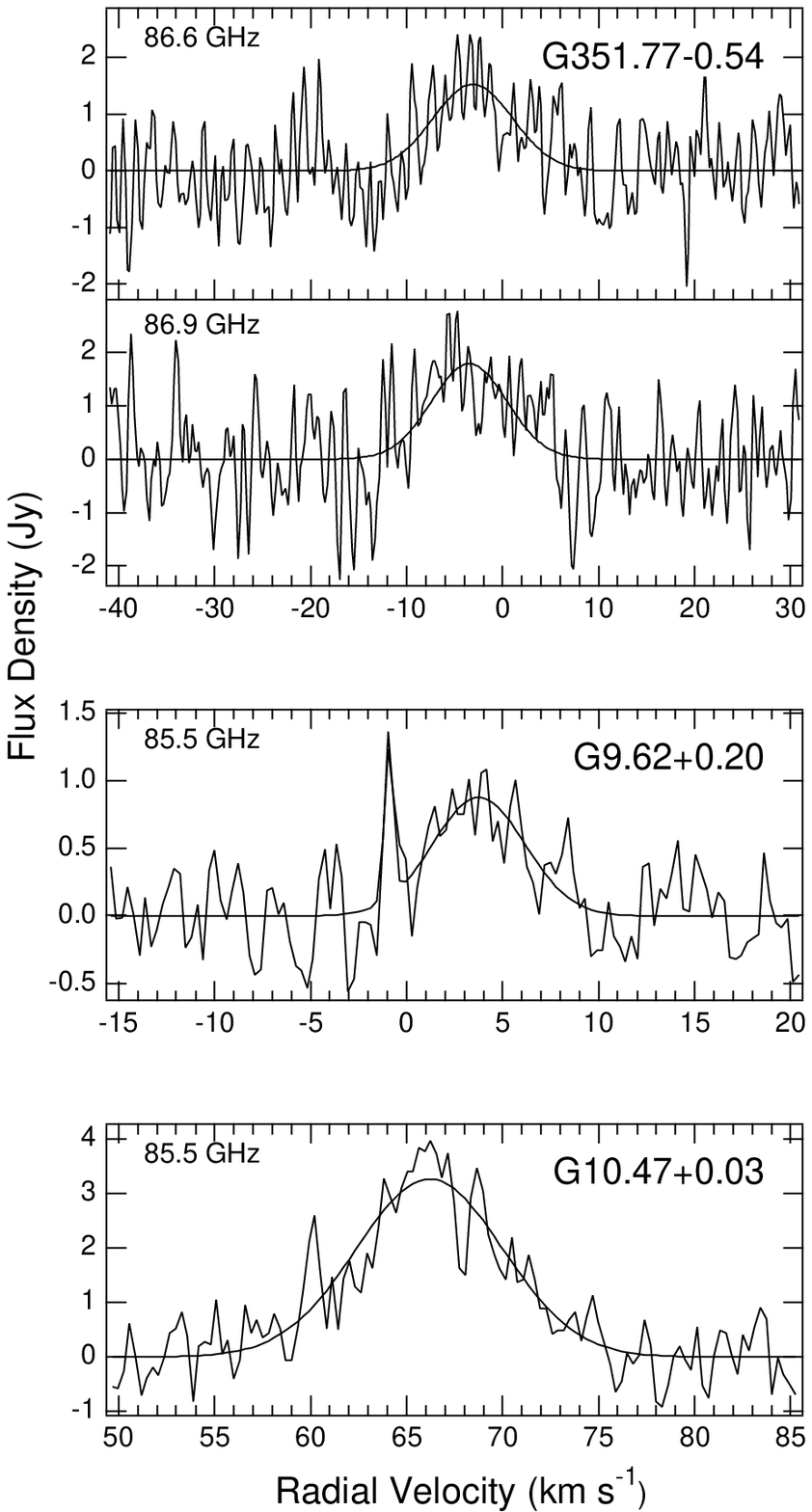}
\caption{Methanol spectra observed at SEST (85.5~GHz) and Mopra (86.6,
86.9~GHz) for G351.77-0.54, G9.62+0.20 and G10.47+0.03.  Note the compressed velocity scale for G351.77-0.54.}
\label{fig:spectra}
\end{figure}

\subsection{Orion KL}

Strong methanol emission was detected at all four frequencies observed in
Orion~KL (Fig.~\ref{fig:orion}) at flux density levels exceeding any of
the maser sources.  The lines have all been previously observed (Lovas
\etal\ 1976, 1982, Hollis \etal\ 1983, Johansson \etal\ 1984, Turner
1991) and are attributed to thermal emission at rotational temperature
100-200~K.  

The general shape of the Orion KL spectra in the four transitions is
similar to the 107- and 108-GHz methanol spectra (Val'tts et al. 1999) but
weaker.  Val'tts et al. interpret their 107- and 108-GHz spectra as
thermal, but with broad and narrow components, both at a velocity of
approximately 7.5 kms.  Our spectra show less evidence for any narrow
component.  

\subsection{G345.01+1.79}

Narrow methanol lines were observed at 85.5, 86.6 and 86.9~GHz near
velocity -22~\kms\ in G345.01+1.79 (Fig.~\ref{fig:g345}).  These coincide
with strong maser peaks at -21.8~\kms\ in the 6.6- and 12.1-GHz spectra
(but note an even stronger 6.6-GHz peak at -17.9~\kms).  Previous
methanol observations of this source at higher frequencies display narrow
maser emission near -22~\kms\ accompanied by (but clearly distinguishable
from) broad thermal emission centred near \mbox{-14~\kms}.  Such behaviour is
seen at 156.6 and 107.0~GHz (Caswell \etal\ 2000), at 108.8~GHz (Val'tts
\etal\ 1999) and in the 157-GHz series (Slysh \etal\ 1995), and is also illustrated in Fig.~\ref{fig:g345} for the 108.8-GHz emission.  The thermal emission is from a site north of the maser, but still within the telescope beam (Caswell et al. 2000).  It seems likely that the 85.5-, 86.6- and 86.9-GHz methanol transitions are exhibiting weak maser emission, while no corresponding thermal component was detected.

\subsection{NGC 6334F}

Methanol emission was detected at five frequencies in NGC~6334F
(G351.42+0.64), as shown in Fig.~\ref{fig:ngc}.  The lines are broad, with no narrow features of distinctive maser character.  An earlier detection of the 86.6- and 86.9-GHz methanol lines at low velocity resolution using the IRAM 30-m telescope (Bachiller \& Cernicharo, 1990) is consistent with our 3-Jy detections if the source is small with respect to the 25\arcsec\ IRAM beam, e.g. $<6$\arcsec\ as found for the 156.6-GHz thermal line by Caswell et al. (2000).  

Maser lines in this source typically appear at velocity near -10 \kms\
(Caswell \etal\ 1995a,b; 2000; Menten \& Batrla 1989, Haschick \etal\
1989), while thermal emission is slightly offset near -8 \kms\ (Caswell
\etal\ 2000, Slysh \etal\ 1999), as illustrated in Fig.~\ref{fig:ngc} for the 6.6-GHz maser emission and the 85.5-GHz thermal emission.  In
our new observations all the methanol lines are near -8 \kms, and so all
lines are probably of thermal origin.  Because of the overlap between the thermal and maser velocity ranges, we use the peak intensity of the thermal line as an upper limit on the maser flux in the modelling which follows.

\subsection{G351.77-0.54}

Weak broad methanol emission near -3~\kms\ was detected at 86.6 and
86.9~GHz in G351.77-0.54 (Fig.~\ref{fig:spectra}), and is likely to be of
thermal origin.  The survey of Caswell \etal\ (2000) found similar broad
emission at 156.6 and 107.0~GHz centred on -2.9~\kms\ in this source,
which was the second strongest thermal source in their sample.

\subsection{G9.62+0.20}

The 85.5-GHz spectrum of G9.62+0.20 shows a broad thermal component centred at 3.8 \kms\ and a narrow spike at -0.9 \kms, possibly of maser origin (Fig.~\ref{fig:spectra}).  Note that the spectrum of the 37.7-GHz methanol line (belonging to the same $J_{-2}-(J+1)_{-1}~E$ line series) contains a maser component (Haschick \etal\ 1989) with velocity and width coinciding with that of the 85.5-GHz spike (to within 0.1 \kms, which is less than the uncertainty 0.3 \kms\ in the rest frequency measurements).  Maser components with similar velocity are also present at 12.1- and 6.6-GHz (Caswell \etal\ 1995a,b).  Although our 85.5-GHz spectrum is very noisy, the spike remains after substantial smoothing, and we regard it as a tentative detection of a maser component at the 3.6 sigma level.  For comparison, the 107.0-GHz spectrum of this source displays two narrow spikes of maser emission at -0.5 and 1.2 \kms\ with a broad underlying thermal component centred at 3.6 \kms\ (Caswell \etal\ 2000).

\subsection{G10.47+0.03}

Methanol emission at 3.3 Jy was detected at 85.5~GHz in G10.47+0.03 (W31), as shown in Fig.~\ref{fig:spectra}.  The broad line centred near 66~\kms\ is at the velocity typical of thermal emission for this source (see e.g. Caswell \etal\ 2000) and we conclude that the line is probably of thermal origin.

\section{Multi-transition maser modelling}

\subsection{The maser model}

We employ the class~II methanol maser model introduced by Sobolev \& Deguchi (1994a) (hereafter the SD model).  Methanol molecules are pumped by infrared radiation from warm dust to their second and first torsionally excited states, from which they decay, giving rise to population inversions between certain levels of adjacent $K-$ladders in the torsional ground state.  The extreme brightness temperatures displayed by some 6.6- and 12.1-GHz methanol masers are reproduced when the model includes amplification of free-free continuum radiation from an underlying ultra-compact \HII\ region, and when the spherical LVG radiative transfer approximation includes a \lq beaming factor\rq\ which distinguishes the radial and tangential optical depths, to better represent an elongated maser geometry.  Details of the pumping cycles are described in Sobolev \& Deguchi (1994b).  Model equations are summarized in the Appendix to Sobolev \etal\ (1997a), in which the SD model was first applied to $A-$species methanol.  Maser predictions for many transitions under a range of representative model conditions appear in SCG97.

The SD model contains three sources of continuum radiation:  the infrared pumping radiation parameterized by \Td, the cosmic microwave background, and the \HII\ region free-free spectrum.  The \HII\ region emits mostly at low frequencies and provides a source of background photons for amplification by the low frequency masers.  The \HII\ spectrum also influences the maser excitation, tending to counteract the IR maser pumping induced by the warm dust, via an increase in the number of low frequency photons penetrating the maser formation region.  We vary only the geometrical dilution factor \WHII, and use a fixed value for the turnover frequency of the \HII\ spectrum (12~GHz), and the electron temperature (the value of 18000~K used in the original SD model gives results very close to those obtained with the canonical value 10000~K).  Radiative transfer is handled by the LVG approximation, with beaming factor \beam\ defined as the ratio of optical depths parallel and perpendicular to the line of sight (Castor 1970).  The tangential methanol column density $N_{M}$ and the line width $\Delta V$ combine to give the parameter \scd, equivalent to the methanol volume density divided by velocity gradient.  Maser brightness temperatures are largely determined by the optical depth along the line of sight, which is proportional to the product of \scd\ and the beaming factor \beam.  

The methanol excitation is also influenced by collisions with molecular hydrogen, parameterized by the gas temperature \Tk\ and volume density \nH.  Unfortunately there are no accurately calculated state-to-state collisional excitation rates for methanol, and the empirical values we adopt (Peng \& Whiteoak 1993) are based on a very small number of measurements.  For comparison some calculations employ nonselective collisions, in which all downward rates between sublevels are assumed equal (scaled to give the same total cross-section).

Statistical equilibrium calculations in the SD model were carried out for a total of 1131 $A-$ and $E-$species methanol energy levels up to $J=18$ for $\vt=0,1,2$  (Mekhtiev, Godfrey \& Hougen 1999).  Masers appear in many transitions over a wide range of values of each of the six model parameters described above, as sampled in SCG97.  The parameter space is too vast to permit a full numerical exploration, so we identify trends by varying model parameters one or two at a time.  Most results are based on an exploration of the cool gas regime with $\Tk=30$~K, typical of rotational temperatures derived from thermal methanol observations (Slysh \etal\ 1999).  We adopt a dust temperature $\Td=175$~K which is sufficient to turn on the $\vt=2$ maser pumping, and within the range of temperatures identified for this type of source (Faison \etal\ 1998).  Beaming factor $\beam=10$ and \HII\ region dilution factor $\WHII=2\times 10^{-3}$ are chosen to generate sufficient brightness temperature at 6.6 and 12.1~GHz to match VLBI observations ($>10^{10}$~K).  The calculations span a range of parameters $\nH=10^{4}-10^{9}$~\cc\ and $\scd=10^{11}-10^{13}$~\ccs.  We have done additional less extensive calculations with $\beam=1,3,20,100$, $\WHII=10^{-2}-10^{-5}$, $\Td=150-250$~K, and $\Tk=30-250$~K.  In general terms the presence or absence of individual masers in the model is governed by the values of the collisional parameters \nH\ and \Tk, while their brightness is determined by the radiative parameters \Td, \scd, \beam\ and \WHII.

\begin{figure*}
\centerline{\epsfxsize=17.5cm\epsfbox{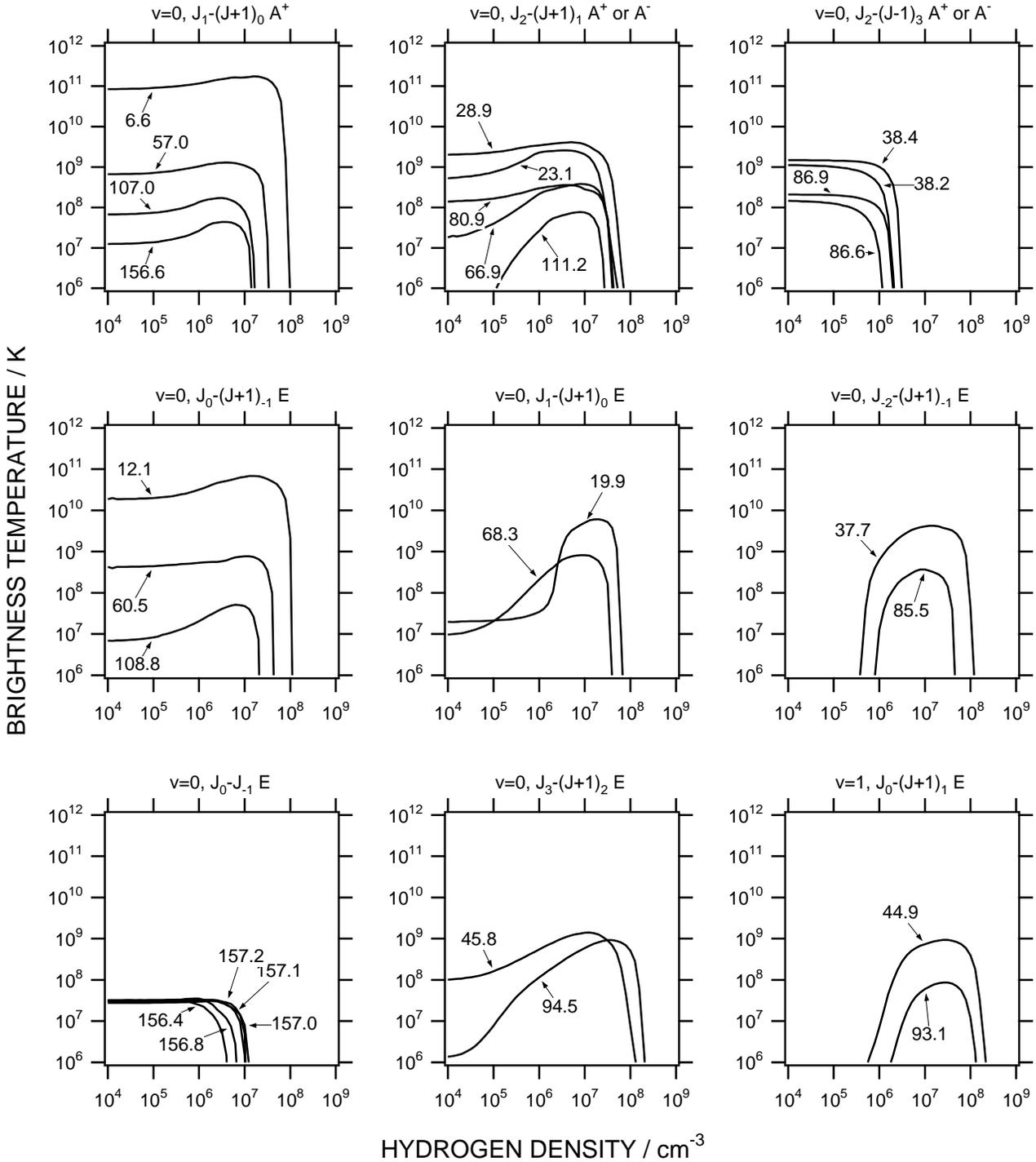}}
\caption{Density dependence of selected class~II methanol masers.  Transitions are grouped into series and labelled with the line frequency in GHz, but for clarity not all members of each series are shown.  The model parameters were as follows:  kinetic temperature $\Tk=30$~K, dust continuum temperature $\Td=175$~K, $\scd=10^{12.3}$ \ccs, beaming parameter $\beam=10$, \HII\ region dilution factor $\WHII=2\times 10^{-3}$.}
\label{fig:density}
\end{figure*}

\begin{figure*}
\centerline{\epsfxsize=17.5cm\epsfbox{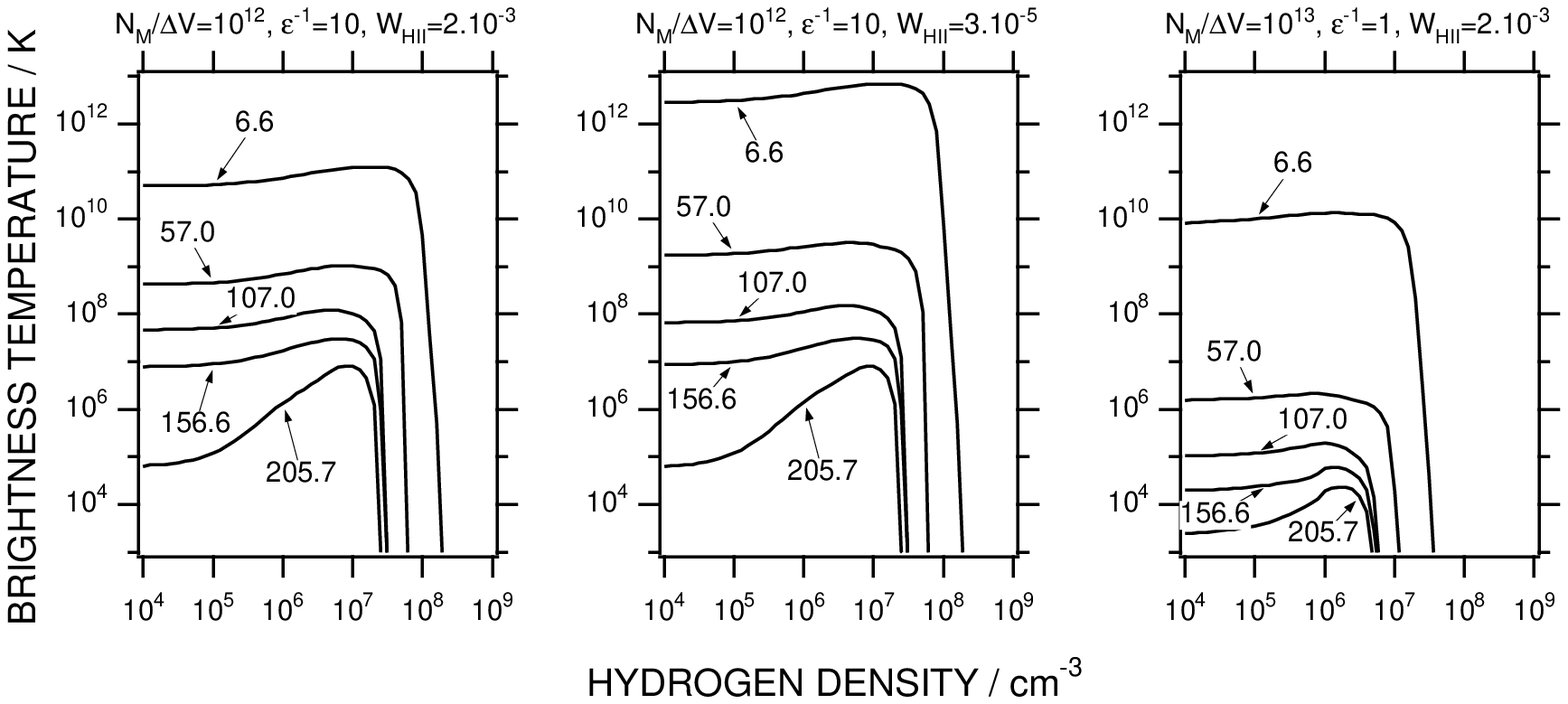}}
\caption{Density dependence of the $\vt=0, J_{1}-(J+1)_{0} \, A^{+}$ maser series for a model with $\beam=10$ (left panel), representing a beamed or elongated geometry.  The centre panel shows the effects of a very distant \HII\ region, while the right panel shows the effects of removing the beaming factor (spherical geometry).  All calculations have $\Tk=30$~K, $\Td=175$~K.  }
\label{fig:ngc_density}
\end{figure*}

The class~II methanol masers are all $b-$type transitions between states of adjacent $K$ quantum number.  Such transitions can be usefully grouped into series involving the same values of $K$ but increasing or decreasing values of $J$.  In the SD model many members of a transition series become masers under the same model conditions, consistent with the suggestion that some methanol $K-$ladders become overpopulated with respect to their neighbours (Wilson \etal\ 1985).

The behaviour of some maser series is illustrated in Fig.~\ref{fig:density}, where we present the model dependence of maser brightness temperature on hydrogen density \nH\ for selected methanol transitions in the cool gas regime.  Many of the masers are present in the low density limit.  Initial density independence is usually followed by growth of maser brightness, and eventual quenching as the density is progressively increased.  No methanol masers are found at densities exceeding $10^{8}$ \cc.  In the low density regime, which extends to approximately $10^{5}$ \cc, the masers are generated by radiative processes and the collisions have very little influence.  The collisions are important in determining the peak maser brightness temperatures in the mid-density regime, particularly for those masers which are not present at low densities (e.g. 85.5 and 93.1~GHz).  They also govern the point at which individual masers \lq switch off\rq\, and hence the extent to which different masers coexist in the model.

Some qualitative features of class~II maser observations are in accordance with Fig.~\ref{fig:density}.  The ubiquitous nature of the 6.6-GHz maser is consistent with its extreme brightness over a wide range of model conditions, while the less prevalent 107.0- and 156.6-GHz masers are weaker by orders of magnitude and have more stringent requirements for strong emission.  In both modelling and observations the 12.1-GHz maser is generally somewhat weaker than the 6.6-GHz maser.  The same applies to lower $J$ members of the corresponding series, with the 108.8-GHz maser likewise weaker than its 107.0-GHz counterpart.  

Fig.~\ref{fig:ngc_density} illustrates the influence of the \HII\ region background continuum for two values of the geometric dilution parameter \WHII.  The \HII\ region brightness temperature falls away sharply above 12 GHz, so it is not directly involved in the maser pumping, and its influence is mainly confined to the 6.6- and 12.1-GHz masers.  These two transitions exhibit the strongest inversions of all the methanol maser lines, and are the first to undergo saturation.  A reduction in \WHII\ (i.e. a more distant or smaller \HII\ region) reduces the maser saturation by reducing the amount of low frequency radiation entering the maser region, while still providing a source of photons for amplification along the line of sight.  The result is that the lowest frequency masers are greatly enhanced, while at millimetre wavelengths the effects are small.

Fig.~\ref{fig:ngc_density} also shows the effects of changing the beaming factor \beam\ from the case where the maser region is optically elongated towards the observer to the optically spherical case where the optical depth is independent of angle.  The line-of-sight optical depth scaling factor \beam\scd\ has the same value for all models displayed in Fig.~\ref{fig:ngc_density}, and these models therefore have the same brightness in the thermal lines (i.e. those with excitation corresponding to LTE).  For static clouds that would correspond to the same line-of-sight methanol column density.  The beaming factor $\beam=10$ represents an optically elongated maser region, in which the radial optical depth (along the line of sight) exceeds the tangential optical depth by a factor of 10.  In this case the maser pumping is very effective, because much of the incident infrared radiation encounters the lower tangential optical depth, allowing many pumping photons to penetrate the maser formation region.  In comparison, in the optically spherical case with beaming factor $\beam=1$ (with a corresponding increase by a factor of 10 in \scd\ to preserve the thermal line brightness), the optical depth encountered by the pumping photons is equally large in all directions.  As shown in Fig.~\ref{fig:ngc_density} the resulting masers are much weaker, particularly at higher frequencies.  This is attributed to the larger tangential optical depth in the spherical case, which limits the infrared pumping.

\subsection{Comparison of model with observations}

A difficulty in comparing models (with predicted brightness \Tb) with observations (with measured flux density) is in converting flux density measurements into maser brightness temperatures, since the required source angular size is too small to be measured with a single dish.  We assume that the masers at different frequencies derive from a common area which is much smaller than the telescope beam, so their relative flux densities can be converted to relative brightness temperatures irrespective of the actual maser spot size.  There is evidence that the class~II methanol masers at different frequencies are, at least, spatially coincident.  The Menten \etal\ (1992) VLBI observations of the methanol masers in W3(OH) have absolute positional information for both the 6.6- and 12.1-GHz spots, and the strongest features coincide in both position and velocity.  ATCA mapping of the 6.6- and 12.1-GHz masers by Norris \etal\ (1993) found an unambiguous coincidence of maser features within 0.02\arcsec\ in several sources.  Low resolution interferometric observations of W3(OH) at 23.1~GHz (Menten \etal\ 1988) and Cep A at 107.0~GHz (Mehringer, Zhou \& Dickel 1997) demonstrate that these sources are small and have counterparts at 12.1~GHz.  The 107.0-GHz masers were found to be coincident with their 6.6-GHz counterparts to within typical measurement uncertainties of 5\arcsec\ for  8 sources checked by Caswell \etal\ (2000).

Under the assumption that the masers are coextensive, the observed flux densities $S(\nu)$ lead to the brightness temperature ratio $R_{obs}=[S(\nu)/S(\nu_0)](\nu_0/\nu)^2$ with respect to a reference line at frequency $\nu_0$.  This may be compared with the model brightness temperature ratio $R_{calc}=\Tb(\nu)/\Tb(\nu_0)$.  The advantage of comparing such ratios is that $\Tb(\nu_0)$ may range over orders of magnitude between different models, in each of which a different source size is implicit.  Brightness temperatures are constrained where VLBI data is available.

There are many reasons why a comparison between the model and a diverse collection of single dish observations can be at best semi-quantitative:  the size of the maser region is unknown, the masers at different frequencies may not be coincident, they may differ in extent, they may be variable, the observed positions are not always identical, there may be significant calibration discrepancies, etc.  We estimate that values of $R_{obs}$ (which range over 5 orders of magnitude) may be uncertain by up to a factor of 3.  Furthermore the current LVG model is likely to be deficient in reproducing details of the spectrum and distribution of the pumping dust, source structure, possible maser saturation effects, etc., and although radiative processes dominate the excitation, many quantitative details at high densities depend also on the poorly known collisional excitation rates.  In view of such uncertainties, order-of-magnitude agreement between $R_{obs}$ and $R_{calc}$ may be all that can be achieved.  Nevertheless, the maser brightness temperatures are so sensitive to the model parameters (Figs.~\ref{fig:density},\ref{fig:ngc_density}) that estimates of the hydrogen density and methanol column density can be obtained.

\subsection{G345.01+1.79}

Narrow lines at velocity near -21.8 \kms\ have been observed in G345.01+1.79 at 13 methanol frequencies, as summarised in Table~\ref{tab:g345}.  All these transitions become class~II masers under some circumstances in the SD model (SCG97).  This source is unusual in that the flux density of the 12.1-GHz maser at -21.8 \kms\ is not much less than that of the 6.6-GHz maser (Caswell \etal\ 1995b).  Furthermore it has the brightest masers yet detected at 107.0,  156.6~GHz (Caswell \etal\ 2000) and the 157-GHz series (Slysh \etal\ 1995), and is the only source in which narrow lines have been detected to date at 85.5, 86.6, 86.9 and 108.8~GHz.  Upper limits for methanol emission at 93.1 and 94.5~GHz are also tabulated.

In the following analysis we explore the ability of the SD methanol maser model to reproduce observational properties of the G345.01+1.79 masers.  We assume that all 13 of the lines are of maser origin, and that they are formed in the same volume of gas.   VLBI at 6.6~GHz has shown the maser spot size to be $<0.0036$\arcsec\ with corresponding brightness temperature $\Tb(6.6) >10^{11.5}$~K (Phillips 1998), while at 12.1~GHz the spot size is $<0.0007$\arcsec\ and $\Tb(12.1) > 10^{11.0}$~K (Ellingsen \etal\ 1998).  There is no mapping data available for the remaining transitions in G345.01+1.79, and evidence of small source size is required to fully establish their maser character, but similarities in maser line profiles suggest that the assumption of coincidence is reasonable.  The masers in Table~\ref{tab:g345} are ranked by their brightness temperatures relative to that of the 6.6-GHz line.  If the 6.6-GHz maser has brightness temperature $>10^{11.5}$~K as suggested by the VLBI observations, then under these assumptions the weakest masers detected have brightness $>10^{7}$~K.

\begin{table*}

\caption{Comparison of observations and modelling for class~II methanol masers in G345.01+1.79.  \Tb(6.6) is the model brightness temperature of the 6.6-GHz maser, and model parameters are defined in Section 4.1.  The relative brightness $R_{obs}$ is the ratio of observed $(S/\nu^{2})$ to its value for the 6.6-GHz line, while for the models we tabulate $R_{calc}=\Tb(\nu)/\Tb(6.6)$ for comparison (see text Section 4.2).  Model~E uses non-selective collisions.}

\label{tab:g345}

\centerline{
\begin{tabular}{clrrrrrrrrr} \hline
&&&&&& \multicolumn{1}{c}{Model A} & \multicolumn{1}{c}{Model B} & \multicolumn{1}{c}{Model C} & \multicolumn{1}{c}{Model D} & \multicolumn{1}{c}{Model E} \\
&&&&& \multicolumn{1}{r}{         log(\nH) } &     4.0 &      6.0 &      5.8 &      7.0 &      6.5 \\
&&&&& \multicolumn{1}{r}{              \Tk } &      30 &       30 &       30 &      100 &       30 \\
&&&&& \multicolumn{1}{r}{              \Td } &     175 &      175 &      175 &      175 &      175 \\
&&&&& \multicolumn{1}{r}{            \beam } &      10 &       10 &       10 &       10 &       10 \\
&&&&& \multicolumn{1}{r}{            \WHII } &  2.E-03 &   2.E-03 &   2.E-03 &   2.E-03 &   2.E-03 \\
&&&&& \multicolumn{1}{r}{        log(\scd) } &    12.3 &     12.3 &     11.8 &     12.2 &     12.0 \\
&&&&& \multicolumn{1}{r}{         \Tb(6.6) } & 8.4E+10 &  1.4E+11 &  4.4E+10 &  6.0E+10 &  6.7E+10 \\
&&&&&&&&&& \\
\vt & Transition & \multicolumn{1}{c}{$\nu$} & \multicolumn{1}{c}{$S$} & $R_{obs}$ && $R_{calc}$ & $R_{calc}$ & $R_{calc}$ & $R_{calc}$ & $R_{calc}$  \\ 
&& \multicolumn{1}{c}{(GHz)} & \multicolumn{1}{c}{(Jy)} &&&&&&& \\ \hline
0 & $5_{1}-6_{0}\,A^{+}$  &   6.6 &  338$^a$ &  1.0       &&   1.0     & 1.0     & 1.0     & 1.0     & 1.0      \\
0 & $2_{0}-3_{-1}\,E$     &  12.1 &  310$^b$ &  2.7E-01   &&   2.2E-01 & 2.7E-01 & 1.7E-01 & 2.4E-01 & 5.4E-01  \\
0 & $3_{1}-4_{0}\,A^{+}$  & 107.0 &   82$^c$ &  9.4E-04   &&   7.9E-04 & 1.2E-03 & 1.2E-03 & 9.9E-04 & 9.6E-04  \\
0 & $5_{0}-5_{-1}\,E$     & 157.1 & 71.1$^d$ &  3.8E-04   &&   3.6E-04 & 2.6E-04 & 2.9E-04 & 3.5E-04 & 4.1E-04  \\
0 & $8_{0}-8_{-1}\,E$     & 156.4 & 66.1$^d$ &  3.6E-04   &&   3.8E-04 & 2.1E-04 & 2.6E-04 & 3.0E-04 & 8.0E-05  \\
0 & $7_{0}-7_{-1}\,E$     & 156.8 &   59$^d$ &  3.2E-04   &&   3.8E-04 & 3.0E-04 & 3.0E-04 & 3.3E-04 & 2.1E-04  \\
0 & $6_{0}-6_{-1}\,E$     & 157.0 & 55.8$^d$ &  3.0E-04   &&   3.7E-04 & 2.8E-04 & 3.0E-04 & 3.3E-04 & 3.3E-04  \\
0 & $4_{0}-4_{-1}\,E$     & 157.2 & 54.9$^d$ &  2.9E-04   &&   3.3E-04 & 2.5E-04 & 2.7E-04 & 2.4E-04 & 4.4E-04  \\
0 & $6_{-2}-7_{-1}\,E$    &  85.5 &   10$^e$ &  1.8E-04   &&  -3.2E-09 & 1.1E-04 &-4.8E-09 &-4.1E-09 & 2.1E-05  \\
0 & $2_{1}-3_{0}\,A^{+}$  & 156.6 &   18$^c$ &  9.7E-05   &&   1.5E-04 & 2.3E-04 & 2.1E-04 & 1.9E-04 & 1.8E-04  \\
0 & $0_{0}-1_{-1}\,E$     & 108.8 &  7.5$^f$ &  8.3E-05   &&   8.2E-05 & 1.8E-04 & 3.5E-05 & 2.3E-05 & 1.7E-04  \\
0 & $7_{2}-6_{3}\,A^{+}$  &  86.9 &  4.1$^e$ &  7.1E-05   &&   2.5E-03 & 8.1E-04 & 9.7E-05 & 8.9E-05 & 5.3E-05  \\
0 & $7_{2}-6_{3}\,A^{-}$  &  86.6 &  2.8$^e$ &  4.9E-05   &&   1.7E-03 & 6.5E-05 & 2.4E-05 & 9.9E-05 & 3.0E-05  \\
\multicolumn{11}{l}{upper limits:} \\
1 &  $1_{0}-2_{1}\,E$  &  93.1 & $<$3.1$^e$ & $<$4.7E-05  &&    5.8E-08 & 1.2E-06 & 1.7E-08 & 3.3E-08 & 4.0E-05 \\
0 &  $8_{3}-9_{2}\,E$  &  94.5 & $<$2.5$^e$ & $<$3.7E-05  &&    1.6E-05 & 9.9E-04 & 1.8E-06 & 1.7E-06 & 4.0E-07 \\ \hline
\end{tabular}}

$^a$ Caswell \etal\ (1995a), $^b$ Caswell \etal\ (1995b), $^c$ Caswell \etal\ (2000), $^d$ Slysh \etal\ (1995), $^e$ this work, $^f$  Val'tts \etal\ (1999)

\end{table*}

We have compared the ratios $R_{obs}$ with those of the model calculations described in Section 4.1, with warm dust ($\Td=175$~K), cool gas ($\Tk=30$~K), and elongated geometry ($\beam=10$).  The relative brightness of the low and high frequency masers in this source is approximately matched with a geometrically dilute \HII\ region spectrum,  $\WHII=2\times 10^{-3}$, giving $R(107.0) \sim 10^{-3}$.  General agreement for many lines is obtained for $\nH<10^{7}$~\cc\ and $10^{11.5}<\scd <10^{12.5}$~\ccs.  The model brightness temperatures at 6.6 and 12.1~GHz attain the VLBI values at the higher ends of these ranges.  The difficulty of fitting a single component cool gas model to all the observations simultaneously is apparent from the trends in Fig.~\ref{fig:density}.  The 86-GHz lines require relatively low densities, while the 85.5-GHz line requires moderately high density.  There is some overlap between the 85.5- and 86-GHz masers at intermediate density when the methanol column density is large.  However the 94.5-GHz nondetection adds a further strong constraint, as it is difficult to find conditions in which the model predicts that the observed 85.5-GHz maser should be brighter than the undetected 94.5-GHz line.  

The results of some illustrative models are included in Table~\ref{tab:g345}.  Model~A represents the low density temperature-independent regime with $\scd=10^{12.3}$~\ccs.  It accounts for many of the observations, but is too bright at 86~GHz, and predicts absorption rather than maser emission for the 85.5-GHz line.  Whether or not the masers are really coincident, the 85.5-GHz observation is clear evidence for a higher gas density origin for at least some of the masers.  Model~B is an intermediate density model ($\nH=10^{6}$~\cc) which matches the observed maser ratios reasonably well (including the 85.5-GHz line), though still too bright at 86.9~GHz.  However, it predicts maser action at 94.5~GHz comparable with that at 107.0~GHz, contrary to our observed upper limit which is more than an order of magnitude fainter.  Model~C has lower column density of methanol ($\scd=10^{11.8}$~\ccs) and gives a better prediction of the 86-GHz brightness while satisfying both the observed upper limits, but has the 85.5-GHz line in absorption.  The difficulty of reconciling the 85.5-GHz maser observation with the 94.5 nondetection persists over the variety of model conditions which we have investigated, including higher gas kinetic temperatures.  Model~D illustrates how a warmer gas temperature ($\Tk=100$~K), with a slightly higher hydrogen density and methanol column density, gives very similar results to the cool gas Model~C, again accounting well for all observations apart from the 85.5-GHz emission.

When the calculations over a range of \nH\ and \scd\ are repeated with non-selective collision rates, the density at which some masers turn on or off alters by up to a factor of 30.  For example, the 86-GHz masers extend to slightly higher density, while the 94.5-GHz maser is restricted to a lower range of densities than before.  While the same range of \nH\ and \scd\ gives broad agreement with the observations, the different collisional rates lead to changes in the details of conditions under which various masers coexist.  The best agreement with observations is exhibited by Model~E of Table~\ref{tab:g345}, with moderately large values of density ($\nH=10^{6.5}$~\cc and $\scd=10^{12.0}$~\ccs), in which all brightness temperature ratios agree to within an order of magnitude and the upper limits are satisfied. Irrespective of the collision model, to obtain approximate agreement between the maser model and the majority of the G345.01+1.79 maser observations requires parameters in the range $\nH=10^{5.5}-10^{7}$~\cc\ and $\scd=10^{11.5}-10^{12.5}$~\ccs\ with $\beam=10$, while \Tk\ is not unambiguously constrained.  We conclude that more reliable collisional excitation rates are required to establish tighter limits on the model parameters.

\subsection{NGC 6334F}

Methanol masers at 8 frequencies have been identified in NGC~6334F near -10.4 \kms, as summarised in Table~\ref{tab:ngc}.  Identification of weak masers in this source is complicated by overlap with broad thermal emission, and we have taken the thermal flux density as a measure of the upper limit for the maser.  This source was the second strongest 6.6-GHz maser in the 245 reported by Caswell \etal\ (1995a).  VLBI at 6.6~GHz has shown the maser spot size to be \mbox{$<0.015$\arcsec} with corresponding brightness temperature $\Tb(6.6) >10^{10.1}$~K, while at 12.1~GHz the spot size is \mbox{$<0.0025$\arcsec} and $\Tb(12.1) > 10^{10.9}$~K (Ellingsen 1996).  Imaging of the -10.4~\kms\ emission at 6.6 and 12.1~GHz shows that it is in the cluster coincident with the brightest continuum emission.

\begin{table*}

\caption{Comparison of observations and modelling for class~II methanol masers in NGC~6334F.  \Tb(6.6) is the model brightness temperature of the 6.6-GHz maser, and model parameters are defined in Section 4.1.  The relative brightness $R_{obs}$ is the ratio of observed $(S/\nu^{2})$ to its value for the 6.6-GHz line, while for the models we tabulate $R_{calc}=\Tb(\nu)/\Tb(6.6)$ for comparison (see text Section 4.2). }

\label{tab:ngc}

\centerline{
\begin{tabular}{clrrrrrrrr} \hline
&&&&&& \multicolumn{1}{c}{Model F} & \multicolumn{1}{c}{Model G} & \multicolumn{1}{c}{Model H} & \multicolumn{1}{c}{Model I}  \\
&&&&& \multicolumn{1}{r}{         log(\nH) } &      6.0 &     6.2 &     5.5 &     6.0 \\
&&&&& \multicolumn{1}{r}{              \Tk } &       30 &      30 &      30 &      30 \\
&&&&& \multicolumn{1}{r}{              \Td } &      175 &     175 &     175 &     175 \\
&&&&& \multicolumn{1}{r}{            \beam } &       10 &      10 &       1 &       1 \\
&&&&& \multicolumn{1}{r}{            \WHII } &   3.E-05 &  3.E-05 &  2.E-03 &  2.E-03 \\
&&&&& \multicolumn{1}{r}{        log(\scd) } &     12.2 &    11.9 &    13.0 &    12.5 \\
&&&&& \multicolumn{1}{r}{         \Tb(6.6) } &  5.9E+12 & 4.0E+12 & 1.2E+10 & 1.2E+10 \\
&&&&&&&&& \\
\vt & Transition & \multicolumn{1}{c}{$\nu$} & \multicolumn{1}{c}{$S$} & $R_{obs}$ && $R_{calc}$ & $R_{calc}$ & $R_{calc}$ & $R_{calc}$   \\ 
&& \multicolumn{1}{c}{(GHz)} & \multicolumn{1}{c}{(Jy)} &&&&&&\\ \hline
0 & $5_{1}-6_{0}\,A^{+}$   &   6.6 &  3300$^a$ & 1.0       &&   1.0     & 1.0     & 1.0     & 1.0      \\
0 & $2_{0}-3_{-1}\,E$      &  12.1 &  1100$^b$ & 1.0E-01   &&   2.2E-01 & 1.9E-01 & 2.3E-01 & 1.8E-01  \\
0 & $2_{1}-3_{0}\,E$       &  19.9 & 412.7$^c$ & 1.4E-02   &&   3.3E-05 & 4.9E-05 & 2.2E-04 & 2.1E-04  \\
0 & $6_{2}-5_{3}\,A^{-}$   &  38.2 &   387$^d$ & 3.6E-03   &&   5.2E-04 & 5.6E-07 & 1.1E-03 & 1.2E-05  \\
0 & $6_{2}-5_{3}\,A^{+}$   &  38.4 &   165$^d$ & 1.5E-03   &&   3.0E-03 & 3.2E-05 & 2.1E-03 & 6.5E-05  \\
0 & $9_{2}-10_{1}\,A^{+}$  &  23.1 &  51.6$^c$ & 1.3E-03   &&   1.2E-02 & 7.1E-03 & 1.4E-03 & 8.3E-04  \\
0 & $7_{-2}-8_{-1}\,E$     &  37.7 &    57$^d$ & 5.4E-04   &&   3.2E-04 & 1.7E-06 & 7.9E-04 & 2.1E-05  \\
0 & $3_{1}-4_{0}\,A^{+}$   & 107.0 &  14.8$^e$ & 1.7E-05   &&   2.3E-05 & 2.9E-05 & 1.2E-05 & 1.8E-05  \\
\multicolumn{10}{l}{upper limits:} \\
0 & $6_{-2}-7_{-1}\,E$    &   85.5 &  $<$7.7$^f$ & $<$1.4E-05  &&     4.9E-08 &  3.2E-09 &  9.5E-06 &  8.6E-07  \\
1 & $1_{0}-2_{1}\,E$      &   93.1 &  $<$3.4$^f$ & $<$5.3E-06  &&     5.8E-09 &  8.2E-10 &  2.8E-07 &  7.8E-08  \\
0 & $8_{3}-9_{2}\,E$      &   94.5 &  $<$4.3$^f$ & $<$6.5E-06  &&     3.3E-05 &  9.2E-06 &  2.3E-05 &  6.1E-06  \\
0 & $7_{2}-6_{3}\,A^{+}$  &   86.9 &  $<$3.6$^f$ & $<$6.4E-06  &&     4.1E-05 &  1.9E-07 &  4.0E-05 &  6.3E-06  \\
0 & $7_{2}-6_{3}\,A^{-}$  &   86.6 &  $<$2.6$^f$ & $<$4.7E-06  &&     2.7E-06 &  7.7E-09 &  1.8E-05 &  1.5E-06  \\ \hline
\end{tabular}}

$^a$ Caswell \etal\ (1995a), $^b$ Caswell \etal\ (1995b), $^c$ Menten \& Batrla (1989), $^d$ Haschick \etal\ (1989), $^e$ Caswell \etal\ (2000), $^f$ this work

\end{table*}

Some of the differences between the methanol masers in NGC~6334F and G345.01+1.79 may be attributed to confusion by thermal emission, and absence of some key observations, but the prevalence of high frequency masers in G345.01+1.79 and their general absence in NGC~6334F indicates significant differences in the pattern of maser excitation.  The 6.6- and 12.1-GHz masers in NGC~6334F have several times greater flux density than those in G345.01+1.79, while the millimetre wavelength masers are significantly weaker, such that only the 107.0-GHz maser has been detected in NGC~6334F, and this is weaker both in absolute and relative terms.  NGC~6334F harbours 5 masers below 40~GHz which have not been sought in G345.01+1.79.  Three of these (37.7, 38.2, 38.4~GHz) belong to the same transition series as the millimetre masers seen only in G345.01+1.79 (85.5, 86.6, 86.9~GHz respectively).  

In the following analysis we consider how the observational properties of the NGC~6334F masers may be reproduced within the SD model.  The \HII\ region dilution factor \WHII\ is the model parameter with greatest influence on the maser brightness at low frequencies, via saturation effects in the strongest masers (see Fig.~\ref{fig:ngc_density} and Section 4.1).  Choosing $\WHII=3\times 10^{-5}$ reproduces the relative intensities of the 6.6- and 107.0-GHz masers, $R(107.0) \sim 10^{-5}$.  Apart from this smaller value of \WHII, Model~F in Table~\ref{tab:ngc} has parameters in the range found suitable for G345.01+1.79 ($\nH=10^{6.0}$~\cc, $\scd=10^{12.2}$~\ccs, \beam=10).  This accounts fairly well for all the observed masers except for that at 19.9 GHz, but fails to meet the upper limits established for the 86- and 94.5-GHz lines.  Model~G has a smaller methanol column density, and satisfies the upper limits at millimetre wavelengths, while giving poorer agreement with the 20-40~GHz maser observations.

An alternative to the very dilute \HII\ region spectrum, which is equally able to reproduce the trend of reduced maser brightness at higher frequency, is to remove the beaming factor by setting $\beam=1$ (see Fig.~\ref{fig:ngc_density}).  In this case the masers arise in a spherical region, with large column density of methanol, and hence optical depth, both laterally and along the line of sight ($\scd=10^{13}$~\ccs).  Model~H illustrates the agreement with the observed maser brightness ratios, while Model~I again shows that the millimetre wavelength upper limits are better satisfied with a smaller methanol column density.  The reduced beaming factor results in a smaller brightness temperature for the 6.6-GHz maser and a correspondingly larger source size, though still meeting the limits set by VLBI.

\section{DISCUSSION}

In comparing the model with a variety of maser observations in two class~II methanol maser sources we find that even in the best case the model predictions differ from the observations by as much as a factor of 10 for one or more transitions.  We estimate that uncertainties in the measured flux densities may contribute up to a factor of 3 to this error, and the other major factor is likely to be inaccurate collisional excitation rates.  A multi-component model, which relaxes our assumption that all the masers are spatially coincident, could potentially account for discrepancies in lines which are underpredicted, if their emission can be attributed to a region with different physical parameters (e.g. the 19.9-GHz maser in NGC~6334F, and the 85.5-GHz maser in G345.01+1.79).  It will not, however, assist with the overprediction of maser emission at 94.5~GHz, accompanying the maser detected at 85.5~GHz in G345.01+1.79, and this we attribute to the poor collision model.  Despite the imperfect agreement between model and observations, order-of-magnitude estimates can be obtained for the hydrogen density and methanol column density in the regions responsible for the majority of the maser lines.

For both sources investigated, modelling suggests a gas density of approximately $10^6$ \cc.  \scd\ is near $10^{12}$~\ccs\ with beaming factor $\beam=10$, or near $10^{13}$~\ccs\ with $\beam=1$. For a typical maser line width $\Delta V \sim 0.5$ \kms, this implies a radial methanol column density of $5 \times 10^{17} \,\rm cm^{-2}$ in both cases.  This is greater than values determined in the vicinity of class I methanol masers in the star formation regions Sgr\,B2 and DR21 (Mehringer \& Menten 1997, Liechti \& Walmsley, 1997).  Since the two sources studied are both exceptional examples of class~II methanol maser sites (G345.01+1.79 being the strongest source of millimetre methanol masers, and NGC~6334F hosting one of the strongest 6.6-GHz masers), it is reasonable that their methanol column densities represent the upper extreme. 

High resolution observations such as those by Caswell (1997) and  by Phillips \etal\ (1998) show that the maximum extent of the maser spot distribution at a single maser site is approximately \mbox{$\sim$ 6000~AU} ($10^{17}$~cm or 30~milliparsec).  If we assume that we do not lie along a preferred line of sight, then the linear size of the maser cluster gives a reasonable estimate of the size of the masing gas cloud, and hence the path length.  These assumptions suggest that the fractional abundance of methanol, $X_{\rm M}$, in the masing regions examined is $\sim 5 \times 10^{-6}$.  This is a little larger than the highest abundance observed by Mehringer \& Menten (1997) toward quasi-thermal cores in Sgr\,B2 ($4 \times 10^{-6}$), but as the latter is an average over a whole core, there should be regions of higher abundance.  Such high methanol abundances are not predicted by existing chemical models.  However, recent observations of a number of high mass protostars confirm earlier indications that methanol is the second most abundant solid-state molecule after water (see Dartois \etal\ 1999 and references therein).  This supports the hypothesis that the methanol abundance in the gas phase is enhanced by evaporation of methanol-ice from grain-mantles during the passage of a shock wave, and if this can be achieved without the simultaneous injection of gas then $X_{\rm M} \sim 5 \times 10^{-6}$ may be plausible.

Class II methanol masers in the two sources modelled are consistent with maser regions exposed to radiation from warm dust ($\Td=175$~K).  While the masers are generally predicted to be brightest for cool gas ($\Tk=30$~K), higher gas temperatures are not inconsistent with the majority of the observations, and temperatures above the sublimation temperature of methanol ice (80~K) can more readily account for the high methanol abundances required.  There are several possible reasons for the remaining discrepancies between model and observations, e.g. that the model has not been adequately explored in all parameter ranges, or that it is deficient in some important aspect of the treatment of radiative or collisional effects.  In particular, since the collisional excitation rates are not well established, some aspects of the temperature or density dependence of individual masers are likely to be poorly modelled.  Our simple model does not account for the source structure, physical parameter gradients, or details of the dust spectrum.  A further possibility is that our underlying assumption that all the masers are coincident is incorrect.  High resolution observations will help to distinguish these alternatives, as will the observation of further maser lines.  For example the 80.9-GHz $7_{2}-8_{1}\rm{A^{-}}$ transition is predicted to be a significant maser in cool gas models (see SCG97), but not for gas temperatures $\Tk>150$~K.

Since the observations were made at different times, some of the discrepancies between model and observations may be due to maser variability.  Caswell, Vaile \& Ellingsen (1995) found that 12.1-GHz masers were more variable than 6.6-GHz masers in the same source, and that the weaker masers show greater variability.  This was attributed to lower levels of saturation in the 12.1-GHz and weaker masers.  We may therefore expect some variability in the weaker millimetre masers, and there is convincing evidence of this at 107.0~GHz in one source (Caswell \etal\ 2000).  The difficulty in accounting simultaneously for the 85.5- and 94.5-GHz observations in G345.01+1.79 may perhaps be due to changes in the maser output over the intervening 2 years.  The 20-40~GHz maser observations in NGC~6334F were made a decade earlier than the millimetre observations.

The methanol maser spectra in the two sources differ significantly, with the low frequency masers favoured in NGC~6334F while the high frequency masers are enhanced in G345.01+1.79.  Modelling suggests at least two possible explanations: a more spherical geometry for the maser region (beaming factor $\beam=1$), or a more distant \HII\ region (geometric dilution factor $3\times 10^{-5}$) in NGC~6334F.  Since the model brightness temperatures of the masers in these two cases differ by two orders of magnitude, future high resolution observations will help to elucidate the situation.

\section{Conclusion}

We have made the first detections of weak class~II methanol masers at 85.5, 86.6 and 86.9~GHz;  these are predicted by a pumping model involving torsional excitation to $\vt=2$ by infrared radiation from warm dust.  By assuming that the masers at different frequencies are coincident we have made the first attempt at multi-transition maser modelling for methanol.  High resolution observations at many frequencies are required to explore these assumptions.  Further observations of the established masers in the frequency range 20-40~GHz would also be valuable for future studies of this type.

The level population and radiation transport modelling of G345.01+1.79 and NGC~6334F suggests that their methanol masers form under similar conditions of hydrogen density near $10^{6}$ \cc\ and column density of methanol as large as $5 \times 10^{17} \,\rm cm^{-2}$.  The physical parameters implied from the maser modelling of these sources are generally consistent with those determined from observations of massive star formation regions.  The major differences between the maser spectra in the two sources can be attributed to differences in the geometry of the maser region, e.g. beamed or elongated geometry favouring the development of the high frequency masers in G345.01+1.79, while the dominance of the lower frequency masers in NGC~6334F is consistent with a more spherical maser region.

The large number of different methanol maser lines observed in these two sources allows us to make a comparison with the simple SD maser pumping model, and to identify some particular shortcomings.  Among these is the model prediction of significant maser action at 94.5~GHz, exceeding our observed upper limit, accompanying the newly detected 85.5-GHz maser.  We conclude that the 94.5-GHz maser is quenched in the maser sources, but that our current best estimates of collision cross-sections do not adequately describe the excitation of this transition.  The density and temperature determination is imprecise because of such uncertainties.

We have identified the ratio of high and low frequency maser intensities as a key feature distinguishing the two sources studied in detail.  This approach is likely to be useful in the analysis of other methanol maser sources, since there is evidence that the 157-GHz maser series has been detected most readily in sources which are also strong in the 107.0-GHz maser.  Of the 13 sources which harbour masers at 107.0~GHz and were searched at 157~GHz (Val'tts \etal\ 1995, Caswell \etal\ 2000, Slysh \etal\ 1999), the four sources in which 157-GHz masers were identified (Slysh \etal\ 1995) are among five with 107.0-GHz masers exceeding 20~Jy.

A multi-transition study of methanol masers requires many assumptions, and even the relatively simple model adopted here has a vast parameter space which has been only partially explored.  Nevertheless, it is encouraging that semi-quantitative trends can be identified from the data presently available, and an explanation of the intensities of all the class~II methanol masers seems to be possible in principle.

\section*{Acknowledgments}

DMC and PDG acknowledge financial support from the Australian Research Council.  AMS thanks University of Sydney Research Centre for Theoretical Astrophysics for the opportunity to visit Australia and participate in Mopra observations.  AMS and SVS thank the Russian Federal Programme \lq Astronomy' and INTAS for partial financial support of the project.  SPE thanks the Queen's Trust for the computing system used to process the 94.5- and 111.2-GHz data.

\end{document}